\documentclass{PHYEAUTH}
\usepackage{graphicx}
\begin{document}
\begin{frontmatter}
\title{Calculations of Electric Capacitance in Carbon and BN Nanotubes,\\
and Zigzag Nanographite (BN, BCN) Ribbons}
\author[address1,address2]{Kikuo~Harigaya\thanksref{thank1}}

\address[address1]{Nanotechnology Research Institute, AIST, 
Tsukuba 305-8568, Japan}
\address[address2]{Synthetic Nano-Function Materials Project, AIST, 
Tsukuba 305-8568, Japan}
\thanks[thank1]{
Corresponding author. Fax: +81-29-861-5375
E-mail: k.harigaya@aist.go.jp}

\begin{abstract}
Electronic states in nanographite ribbons with zigzag edges
are studied using the extended Hubbard model with nearest 
neighbor Coulomb interactions.  The electronic states with 
the opposite electric charges separated along both edges are 
analogous as nanocondensers.  Therefore, electric capacitance, 
defined using a relation of polarizability, is calculated to 
examine nano-functionalities.  We find that the behavior of the 
capacitance is widely different depending on whether 
the system is in the magnetic or charge polarized phases.  
In the magnetic phase, the capacitance is dominated by 
the presence of the edge states while the ribbon width is small.  
As the ribbon becomes wider, the capacitance remains with 
large magnitudes as the system develops into metallic 
zigzag nanotubes.  It is proportional to the inverse 
of the width, when the system corresponds to the semiconducting 
nanotubes and the system is in the charge polarized phase also.
The latter behavior could be understood by the presence of 
an energy gap for charge excitations.  In the BN (BCN) nanotubes 
and ribbons, the electronic structure is always like of semiconductors.
The calculated capacitance is inversely proportional to 
the distance between the positive and negative electrodes.
\end{abstract}

\end{frontmatter}

\section{Introduction}

Nano-carbon (C) materials and hetero-materials including borons (B)
and nitrogens (N) have been attracting much attention 
both in the fundamental science and in the interests of
application to nanotechnology devices [1,2].  Their physical 
and chemical natures change variously depending on geometries 
[1-3].  In carbon nanotubes, diameters and chiral arrangements 
of hexagonal pattern on tubules decide whether they are 
metallic or not [1,2].

In nanographites, the edge atoms strongly affect the electronic 
states [3], and there are nonbonding molecular orbitals 
localized mainly along the zigzag edges.  
Recently, we have studied the competition between the spin 
and charge orderings due to the on-site and nearest 
neighbor Coulomb interactions [4].  The nearest neighbor 
Coulomb interaction stabilizes a novel charge polarized 
(CP) state with a finite electric dipole moment in 
zigzag ribbons, and it competes with the spin 
polarized (SP) state.  Though it has been discussed that 
the transverse electric field might induce the first order 
phase transition from the SP state to the CP state [4], 
we need further study in order to reveal what roles such the novel
SP and CP states play in actual physical quantities
measurable in experiments.

\vspace{3mm}
\begin{center}
\resizebox{!}{2cm}{\includegraphics{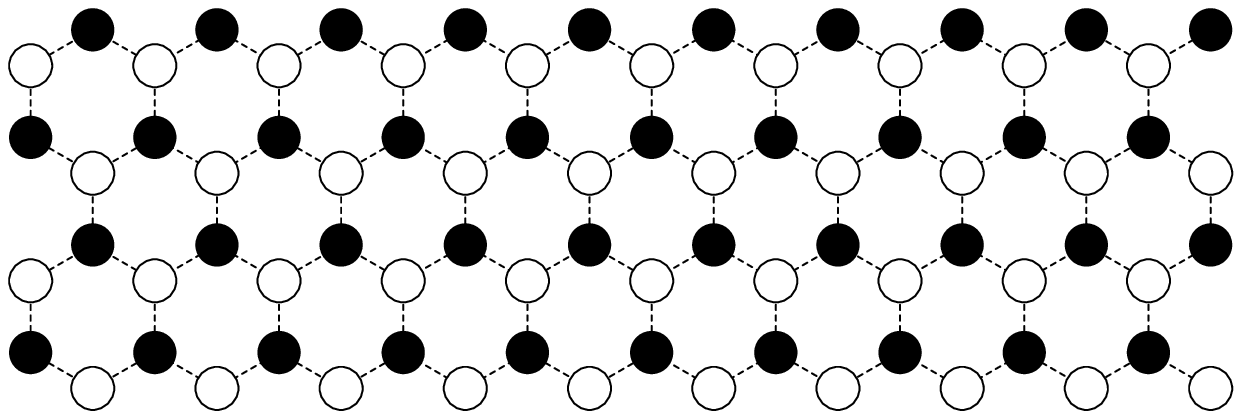}}
\end{center}
\vspace{3mm}
\noindent
{\small Fig. 1. Schematic structure of the bipartite 
ribbon with zigzag edges.  The filled and open circles 
are $A$ and $B$ sites, respectively.}

\section{Carbon nanotubes and nanographite}

Figure 1 shows the schematic structure of the nanographite 
ribbon with zigzag edges.  The ribbon is a system with a 
finite width and an infinite length.  Here, we treat the 
ribbon with a finite width and a finite
length, using the periodic boundary condition for the 
one-dimensional direction.  The lattice sites are classified
as $A$ and $B$ sites due to the bipartite character.
We consider the extended Hubbard model with the inter-site
hopping integral $t$ of $\pi$-electrons, the onsite $U$ and
nearest-neighbor $V$ interactions [4,5].  The typical 
results will be shown for the representative parameters.  
Usually, the relation $U > V$ would be satisfied, and the 
magnitude would be $U \sim$ (a few) eV.  The detailed values 
could be determined by comparing the calculation of the exciton
effects with photophysical experiments, for example.  The relative 
stabilities between the CP and SP states have been 
investigated and summarized in the phase diagram [4].
While $U$ is fairly larger than $V$, the SP state is
more stable.  As $V$ becomes stronger, the system exhibits
the first order phase transition from the SP state
to the CP state.  The charge order occurs in the CP phase, 
while the spin order occurs in the SP phase.

In order to examine nano-functionalities
as``nano-size condensers", electric capacitance of the 
nanographite ribbons is calculated.   We assume 
that the two sets of carbon atoms at the zigzag edges are 
regarded as positive and negative electrodes, respectively.  
The spacer between two electrodes is the inner part of the 
zigzag nanoribbon.  The absolute value of the net variation 
of the accumulated charge is divided by the strength of 
the small applied voltage, and the capacitance is obtained.  
We assume that the bond length between carbons is 1.45~\AA.

Figure 2 shows the calculated results for the system in
the SP state.  The actual magnitudes in the
logarithmic scale are plotted in Fig. 2
(a) against the ribbon width in the scale of~\AA, and 
their inverse values are shown in (b).  When the width
$N$ of the system increases, it develops into the $(L/2,0)$ 
zigzag nanotube due to the periodic boundary condition along 
the zigzag-edge direction.  The nanotube is metallic when $L/2$ is 
a multiple of three, and is semiconducting for others [2] 
in the noninteracting model.  While the 
ribbon width is small (approximately $< 10$~\AA), the 
capacitance would be dominated by the presence of the edge 
states.  Even though the ribbon width becomes larger,
the capacitance remains with large magnitudes for $L=18$,
as shown by squares in Fig. 2 (a).  This would be related
with the resulting metallic properties of the long enough
$(9,0)$ nanotube.  On the other hand, the capacitance for 
$L=20$ is proportional to the inverse width for larger widths,
as shown in Fig. 2 (b).  The presence of the semiconducting
gap in the $(10,0)$ nanotube might result in the distinctive 
difference.

\vspace{3mm}
\begin{center}
\resizebox{!}{6cm}{\includegraphics{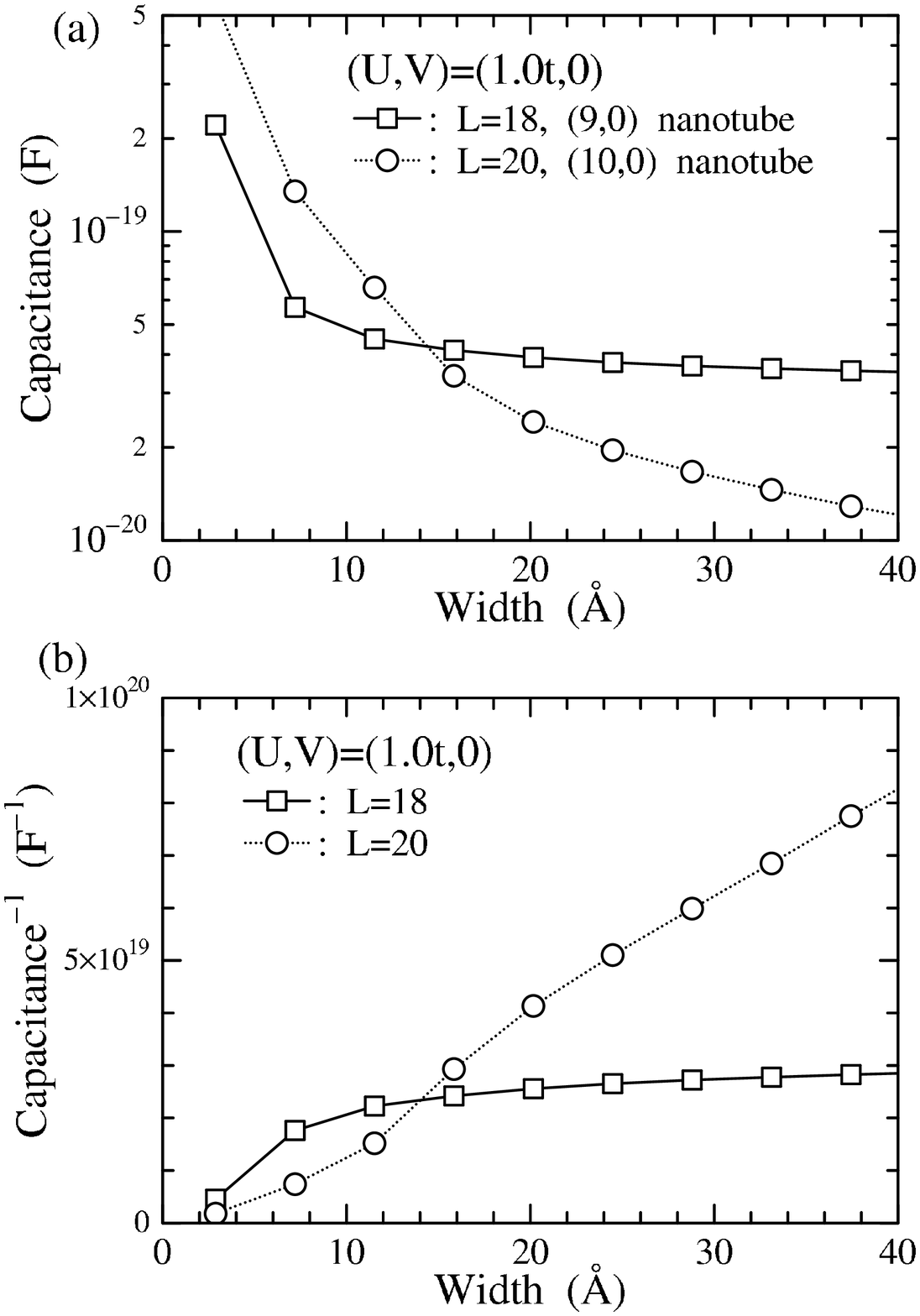}}
\end{center}
\vspace{3mm}
\noindent
{\small Fig. 2.  The electric capacitance calculated for the SP 
state at $(U,V)=(1.0t,0)$. Two sets of the ribbon lengths $L=18$
(squares) and 20 (circles) are considered.  The magnitude of the 
capacitance (a) and its inverse (b) are plotted against the ribbon
width in the scale of~\AA.}
\vspace{3mm}

Figure 3 summaries the calculated capacitance for the 
system in the CP state.  The two sets of $L=18$ and 20
are shown, too.  We find that the capacitance is inversely 
proportional to the graphite width [Fig. 3 (b)].  This 
behavior does not depend on whether the long enough system 
is the metallic (9,0) or semiconducting (10,0) nanotube.  
As the system has the charge orders, electrons would
become less mobile than those in the system without ordered
states.  The behavior could be understood by the presence 
of an energy gap for charge excitations.

\vspace{3mm}
\begin{center}
\resizebox{!}{6cm}{\includegraphics{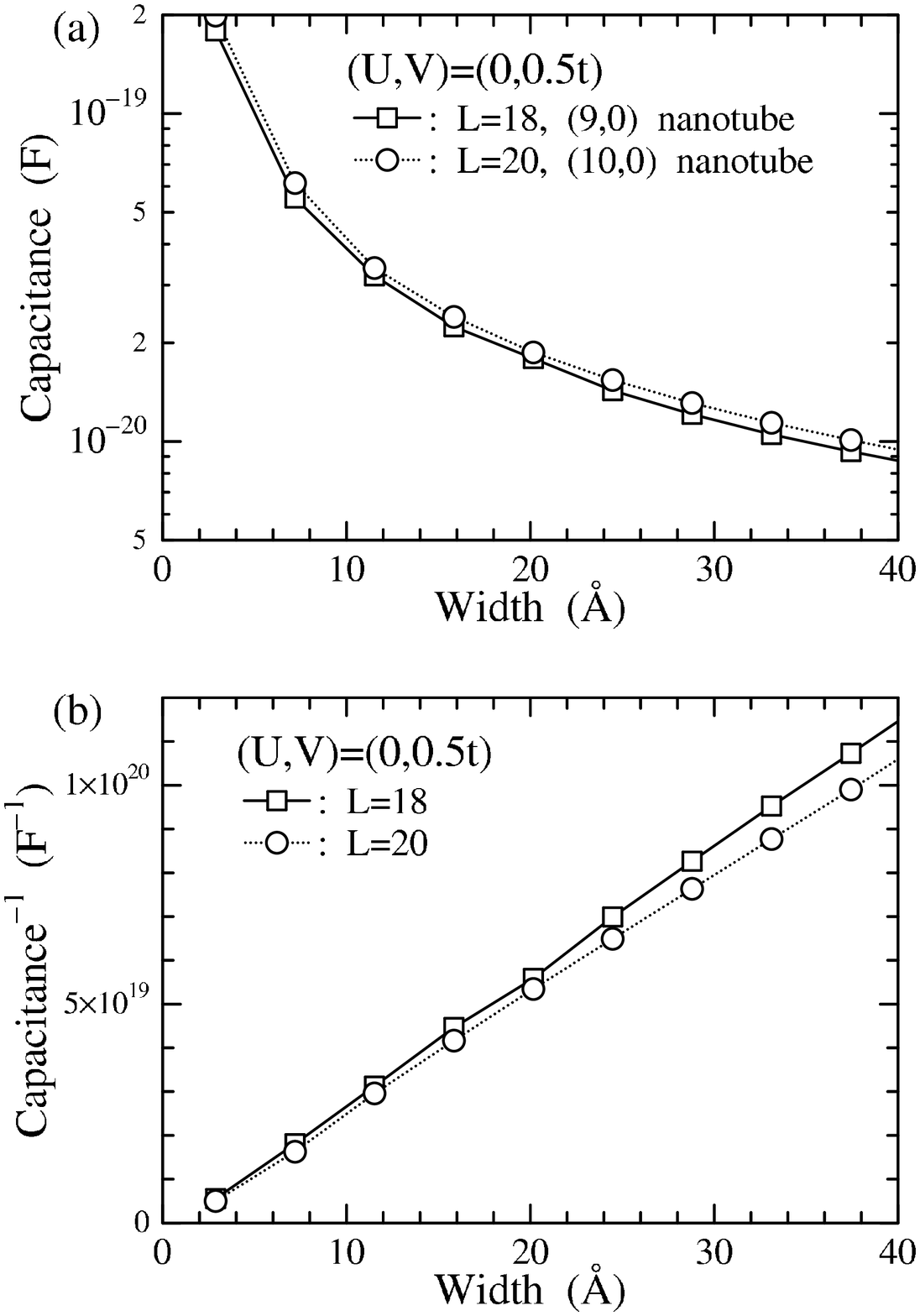}}
\end{center}
\vspace{3mm}
\noindent
{\small Fig. 3. The electric capacitance calculated 
for the CP state at $(U,V)=(0,0.5t)$.  Two sets of the 
ribbon lengths $L=18$ (squares) and 20 (circles) are considered.  
The magnitude of the capacitance (a) and its inverse (b) 
are plotted against the ribbon width in the scale of~\AA.}
\vspace{3mm}

The intrinsic capacitance of the single wall carbon 
nanotube can be estimated as follows.  We consider a
nanotube with finite length which is present in a 
quantum box of the length $L_y$.  The wavenumber
is quantized with the interval $\Delta k = 2\pi/L_y$.
Addition energy of an electron to the nanotube in the
quantum box can be equated with an energy as a dielectric
system.  The addition energy can be estimated to be $\Delta E 
=(h/2\pi) v_F \Delta k \cdot (1/2) \cdot (1/2)$, where 
$v_F$ is the Fermi velocity, the first factor (1/2) comes from 
the degeneracies of spin, and the second one (1/2) is due
to the degeneracy of electronic states near the Fermi
energy.  Here, we assume that the nanotube interacts
with circumstances, and the degeneracies of spin
and electronic states are lifted slightly. Therefore, $\Delta E$
is the overall average level spacing including spin
and number of electronic bands.  By setting 
$\Delta E = e^2/ 2 C_Q L_y$, we obtain the quantum 
capacitance per unit length: $C_Q = 2 e^2/h v_F$.
For carbon nanotubes $v_F = 8 \times 10^5$ (m/s) [6], 
so that $C_Q = 100$ (aF/$\mu$ m) $= 10^{-20}$ (F/\AA).
In fact, the capacitance per unit length has been
measured to be 190 (aF/$\mu$ m) [7] for example,
and the above rough estimation explains the experimental
magnitude fairly well.  Such the order of magnitudes 
of the estimation and experimental value also agrees 
with that of the capacitance obtained in the calculation
within the difference of a few order of magnitudes.
Even though the detailed definitions of the capacitances
are different mutually, the resulting order of magnitudes
should reflect quantum characters of electronic systems.

\vspace{3mm}
\begin{center}
\resizebox{!}{2cm}{\includegraphics{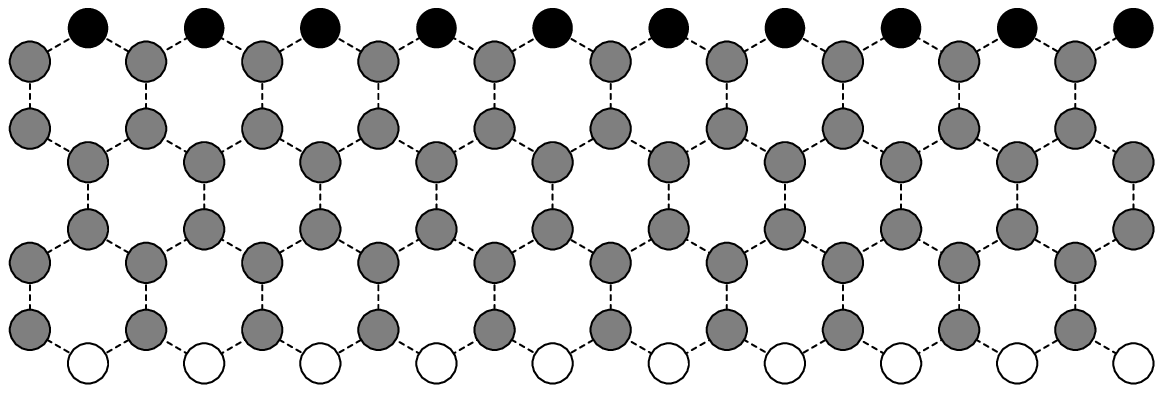}}
\end{center}
\vspace{3mm}
\noindent
{\small Fig. 4. Schematic structure of the BCN
nanoribbon with zigzag edges.  The filled, shaded,
and open circles are B, C, and N atoms, 
respectively.}

\section{BN and BCN systems}

We consider hetero-materials composed of B, C, and N.
One is the BN ribbon with zigzag edges, and its structure
is that of Fig. 1, where B and N atoms exist at the
$A$ and $B$ sites, respectively.  The other is the BCN
system, where one set of edge sites are occupied with B,
and another set of edge sites are with N.  The inner region
of the ribbon is composed of C.  The structure is shown
in Fig. 4.  The experiments of the low concentration limit of 
B and/or N doping into carbon nanotubes have sometimes 
suggested accumulation of impurity atoms at edge sites [8].  
The formation of zigzag nanotubes is favored.  Therefore, 
we choose this structure as a model system.  The site energies 
at the B and N are taken to be $E_{\rm B} = +t$ and $E_{\rm N} 
= -t$.  The total electron number is same with that of the
site number.  Such the strong site energy difference gives rise
to huge charge polarizations.  Therefore, the region
of the CP state extends in the phase diagram, and the SP
state is highly suppressed.  The realistic values of
the interactions correspond to CP states in the phase
diagram.  In the following, we look at representative
behaviors of the electric capacitance of the BN
and BCN systems.

Figure 5 shows the electric capacitance of the zigzag BN
ribbons with changing the ribbon width.  Its raw value (a)
and the inverse (b) is plotted.  We take three parameter
sets of Coulomb interactions.  The system is in the CP 
state for these parameters.  Figure 6 shows the calculated
results for the BCN system.  The system is in the CP state, too.
Both calculations show the almost inversely proportional
behaviors of the capacitance with respect the ribbon
width.  There is a huge electronic gap due to the strong
site energy difference.  The system is a semiconductor
intrinsically.  The strong charge excitation energy gap
results in the inversely proportional behaviors, in contrast
to the saturating behaviors of the SP state found for the
metallic carbon systems (Fig. 2).

\vspace{3mm}
\begin{center}
\resizebox{!}{6cm}{\includegraphics{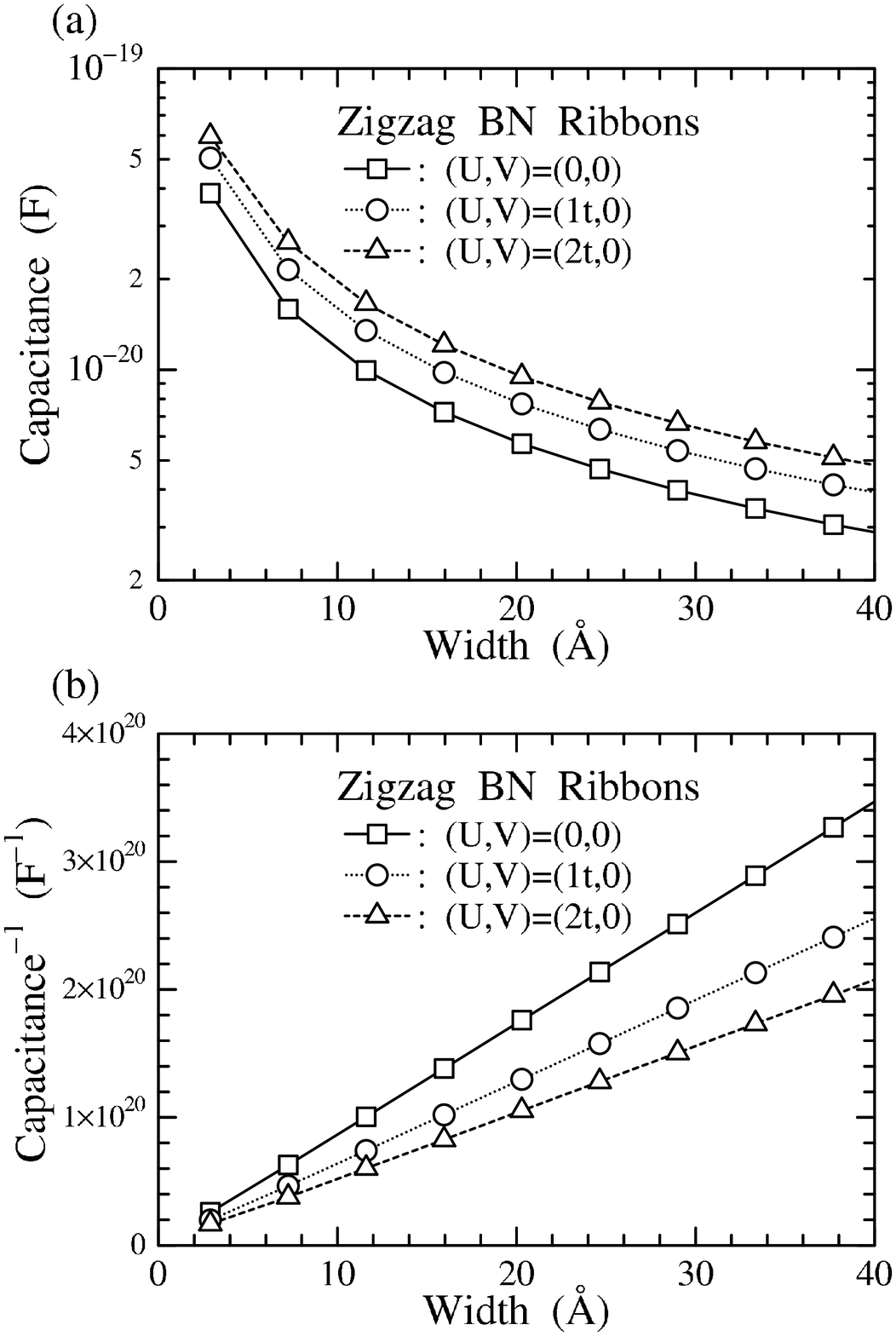}}
\end{center}
\vspace{3mm}
\noindent
{\small Fig. 5. The electric capacitance calculated 
for the CP state of the zigzag BN ribbons at $U = 0$,
$1t$, and $2t$ with $V=0$.  The ribbon length is $L=20$.  
The magnitude of the capacitance (a) and its inverse (b) 
are plotted against the ribbon width in the scale of~\AA.}

\section{Summary}

Electric capacitance has been calculated in order to test 
the nano-functionalities of carbon and BN (BCN) nanotubes 
and ribbons.  In the magnetic phase of the nanographite, 
as the ribbon becomes wider, the capacitance remains with 
large magnitudes as the system develops into metallic zigzag 
nanotubes, while it is proportional to the inverse of the width 
when the system corresponds to the semiconducting 
nanotubes.  In the charge polarized phase, the capacitance 
is inversely proportional to the graphite width. 
In the BN (BCN) nanotubes and ribbons, the electronic 
structure is always like of semiconductors.  The calculated 
capacitance is inversely proportional to the distance 
between the positive and negative electrodes.

\vspace{3mm}
\begin{center}
\resizebox{!}{6cm}{\includegraphics{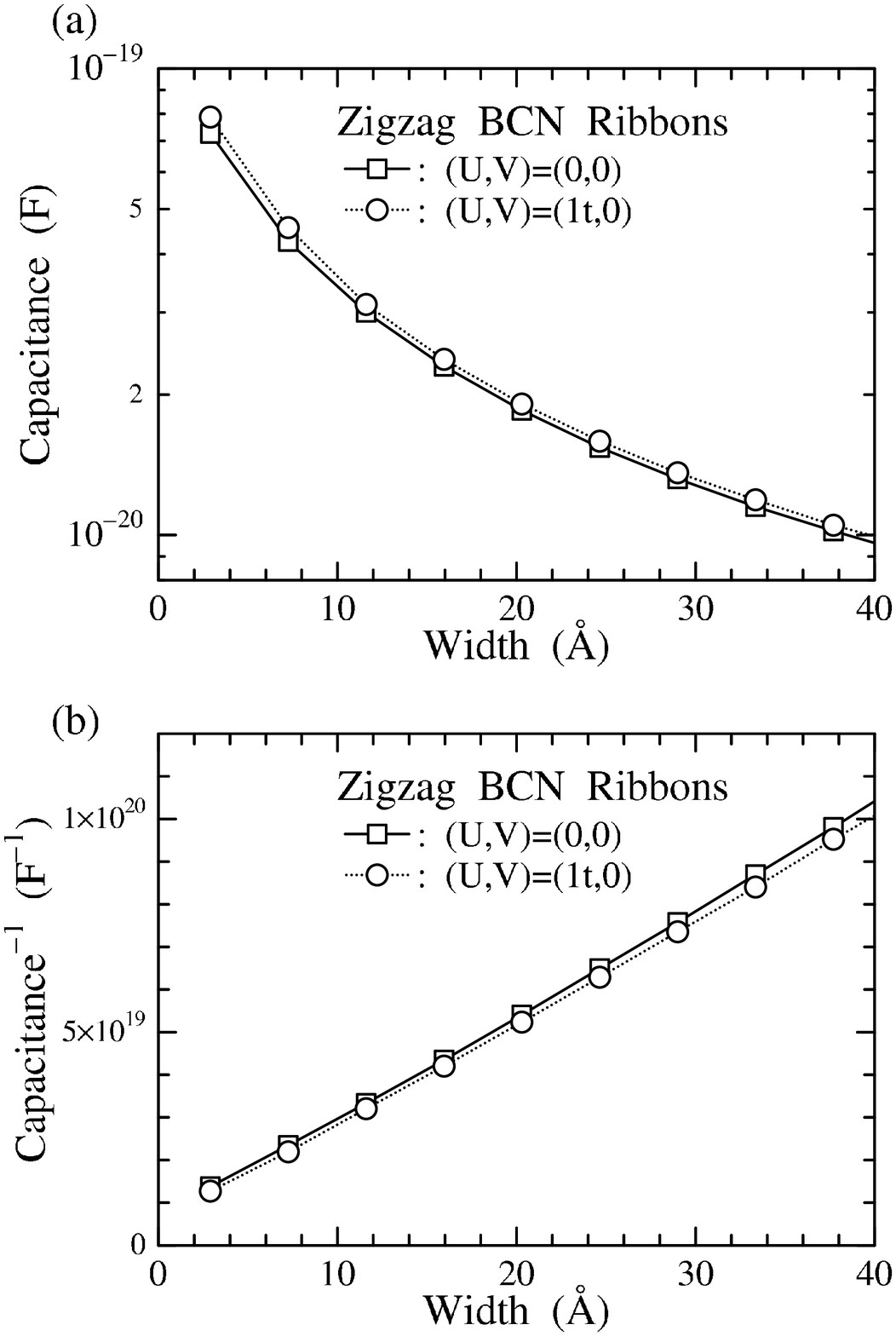}}
\end{center}
\vspace{3mm}
\noindent
{\small Fig. 6. The electric capacitance calculated 
for the CP state of the zigzag BCN ribbons at $U = 0$
and $1t$ with $V=0$.  The ribbon length is $L=20$.  
The magnitude of the capacitance (a) and its inverse (b) 
are plotted against the ribbon width in the scale of~\AA.}

\end{document}